\documentclass[epj]{svjour}

\usepackage{cite,graphicx,amsmath,amsfonts}

\makeatletter

\newcommand{\tsi}{\vec}    
\newcommand{\tsii}{\tens}   
\newcommand{\tsiv}{\tens}   
\newcommand{\dd}{\mathrm{d}} 
\newcommand{\FF}{F}          
\newcommand{\xx}{x}          
\newcommand{\bxx}{\tsi{\xx}} %
\newcommand{\uu}{u}          
\newcommand{\buu}{\tsi{\uu}} %
\newcommand{\ww}{w}          
\newcommand{\bww}{\tsi{\ww}} %
\newcommand{\supu}[1]{#1^{\uu}}  
\newcommand{\supw}[1]{#1^{\ww}}  
\newcommand{\Epar}{E^{\parallel}}
\newcommand{\Eperp}{E^{\perp}}   
\newcommand{\ep}{\varepsilon}    
\newcommand{\bep}{\tsii{\ep}}    %
\newcommand{\epu}{\supu{\ep}}    
\newcommand{\bepu}{\supu{\bep}}  %
\newcommand{\epw}{\supw{\ep}}    
\newcommand{\bepw}{\supw{\bep}}  %
\newcommand{\si}{\sigma}     
\newcommand{\bsi}{\tsii{\si}} %
\newcommand{\CC}{\tsiv{C}}          
\newcommand{\bCC}{\CC} %
\newcommand{\bb}{f}          
\newcommand{\bbb}{\tsi{\bb}} %
\newcommand{\ff}{s}          
\newcommand{\DD}{\tsiv{D}}          
\newcommand{\bDD}{\DD}
\newcommand{\GR}{\tsiv{G}}          
\newcommand{\bGR}{\GR}
\newcommand{\Eins}{\tens{1}} 
\newcommand{\delt}{\delta}   
\newcommand{\RP}{\Gamma}     
\newcommand{\Lapl}{\Delta}   
\newcommand{\abs}[1]{\lvert#1\rvert}  
\@ifundefined{exp}{}{}
\DeclareMathOperator{\Div}{div}   
\DeclareMathOperator{\Grad}{grad} 
\newcommand{\RE}{\mathbb{R}}
\newcommand{\textname}{\textsc}
\renewcommand{\emph}[1]{``#1''}
\newcommand{\dimensional}{-di\-men\-sional}
\newcommand{\fold}{-fold}
\makeatother

\begin{document}

\title{Elastic Green's Function of Icosahedral Quasicrystals}

\author{J. Bachteler \and H.-R. Trebin}
\institute{Institut f\"ur Theoretische und Angewandte Physik\\
  Universit\"at Stuttgart\\
  Pfaffenwaldring 57\\
  D-70569 Stuttgart\\
  Germany}
\date{}

\abstract{
  The elastic theory of quasicrystals considers, in addition to the
  \emph{normal} displacement field, three \emph{phason} degrees of
  freedom.  We present an approximative solution for the elastic
  Green's function of icosahedral quasicrystals, assuming that the
  coupling between the phonons and phasons is small.
}

\maketitle

\section{Introduction} 
\label{Introduction}
The displacement field of an infinite linear elastic medium caused by
external forces can be calculated immediately, when the elastic
Green's function is available. Unfortunately, analytical expressions
for the Green's function are known only for special cases.
\textname{Lord Kelvin}\cite{itapdb:Thomson1848} has provided a solution
for the isotropic medium, \textname{Lifshitz} and
\textname{Rosentsveig}\cite{itapdb:Lifshitz1947} and
\textname{Kr\"oner}\cite{itapdb:Kroner1953} for hexagonal crystals.
Approximations have been presented by \textname{Dederichs} and
\textname{Leibfried}\cite{itapdb:Dederichs1969} for cubic crystals and
by \textname{Kr\"oner}\cite{itapdb:Kroner1953} for arbitrary
anisotropic systems.

In the case of quasicrystals\cite{itapdb:Shechtman1984}, three
additional phason degrees of freedom lead to a more
general theory of elasticity%
\cite{itapdb:Levine1985a,itapdb:Bak1985a,itapdb:Lubensky1985a,%
  itapdb:De1987c,itapdb:Ding1993}. %
The corresponding Green's functions for planar pentagonal (\textname{De} and
\textname{Pelcovits}\cite{itapdb:De1987c}), decagonal and
dodecagonal (\textname{Ding} et al\cite{itapdb:Ding1995a})
quasicrystals are known. However, the solution for icosahedral
quasicrystals with five elastic constants (two for phonons, two for
phasons, and one for the coupling between phonons and phasons) has not
been established yet.

It is the intention of this paper to derive an analytical
approximation of the elastic Green's function for icosahedral
quasicrystals.

The paper is organised as follows. In section \ref{elastic-theory} we
recall the elastic theory of icosahedral quasicrystals.  We present
the Green's function in the case of the \emph{spherical approximation}
and for vanishing coupling between phonons and phasons.  Starting from
this solution and using perturbation theory as well as an invariant
basis for the Green's function tensor, we derive in section
\ref{solution} an approximative solution for finite phonon-phason
interactions. The last section presents the phonon and phason fields
for some special cases.

%
%
%
%
\section{Elastic theory for icosahedral quasicrystals} 
\label{elastic-theory}
\subsection{Equations} 
\label{elastic-theory-equations}

The elastic theory of quasicrystals%
\cite{itapdb:Bak1985a,itapdb:Levine1985a,itapdb:Lubensky1985a,%
  itapdb:Ishii1989e,itapdb:Trebin1991} studies, apart from the
classical \emph{phonon} displacement field $\buu(\bxx)$, also the
\emph{phason} displacement field $\bww(\bxx)$.  Icosahedral
quasicrystals can be described as a three\dimensional{} cut through a
six\dimensional{} cubic crystal.  The six\dimensional{} space divides into
two three\dimensional{} subspaces, which are invariant under icosahedral
symmetry operations, the \emph{physical} or \emph{parallel} cutting
space $\Epar$ and the \emph{perpendicular} space $\Eperp$.  While the
phonon displacement vector $\buu$ is an element of $\Epar$, which
transforms according to the icosahedral irreducible representation
$\RP^{3}$, the phason displacement vector $\bww \in \Eperp$ transforms
according to the other three\dimensional{} icosahedral irreducible
representation $\RP^{3'}$.

Both displacements only depend on the position vector $\bxx \in
\Epar$.  The variations of this displacement fields are described by
the strain fields
\begin{equation}
  \label{eq:definition-strain}
    \epu_{ij} = \frac{1}{2}
       \left(
         \frac{\partial\uu_{i}}{\partial\xx_{j}} 
          + \frac{\partial\uu_{j}}{\partial\xx_{i}} 
       \right)
    \qquad
    \epw_{ij} = \frac{\partial\ww_{i}}{\partial\xx_{j}}
    \qquad
    \bep = 
    \begin{bmatrix}
      \bepu \\ \bepw
    \end{bmatrix}.
\end{equation}
In a linear theory the elastic energy is a quadratic form of the
strain field:
\begin{equation}
  \label{eq:elastic-energy}
  \FF(\bep) = \frac{1}{2}\CC_{\alpha i\beta j}\ep_{\alpha i}\ep_{\beta j} 
    \qquad
  \alpha,\beta = 1 \ldots 6 \qquad i,j = 1 \ldots 3 .
\end{equation}
Differentiation with respect to the strain $\ep_{\alpha i}$ leads to
Hooke's law
\begin{equation}
  \label{eq:Hooke-law}
  \frac{\partial\FF}{\partial\ep_{\alpha i}}=:
    \si_{\alpha i} = \CC_{\alpha i\beta j}\ep_{\beta j},
\end{equation}
where $\bsi$ denotes the generalized stress field.  The meaning of the
\emph{phason stress} $\supw{\bsi}$ is not quite clear.  According to
the definition \eqref{eq:Hooke-law} it is the force conjugated to the
phason strain $\supw{\bep}$.  Phason strain is---in an atomic
description---related to rearrangement of atoms on a small scale
(so-called \emph{flips}).  Forces which inhibit such rearrangements
might be considered as \emph{pinning forces}.

It is shown by group theoretical arguments\cite{itapdb:Levine1985a}
that Hooke's elastic tensor $\bCC$ for icosahedral
symmetry contains only five independent elastic constants
($\mu_{1},\ldots,\mu_{5}$).  From the transformation rules for
$\buu,\bxx: \RP^{3}$ and $\bww: \RP^{3'}$
immediately follows \cite{itapdb:Trebin1991} the transformation behaviour of
the strain fields
\begin{equation}
  \label{eq:trafo-strain}
  \supu{\bep} : \RP^{1}+\RP^{5} \qquad \supw{\bep} : \RP^{4}+\RP^{5},
\end{equation}
which means that they can be written in an icosahedral basis
\begin{equation}
  \label{eq:strain-irreducible}
  \supu{\bep} \rightarrow \bep^{\uu1}+\bep^{\uu5}  \qquad
  \supw{\bep} \rightarrow \bep^{\ww4}+\bep^{\ww5},
\end{equation}
where $\bep^{(\uu/\ww)d}$ is a $d$-dimensional vector transforming
according to $\RP^{d}$ (see appendix \ref{basis-strain} for details).  In
this sym\-me\-try-adjusted coordinate system for the strains $\bep$ and in
an analogous way for the stresses $\bsi$ Hooke's law can be written as
\begin{equation}
  \label{eq:Hooke-law-explicit}
  \begin{bmatrix}
    \bsi^{\uu1} \\ \bsi^{\uu5} \\ \bsi^{\ww4} \\ \bsi^{\ww5}
  \end{bmatrix}
  =
  \begin{bmatrix}
    \mu_{1} & 0 & 0 & 0 \\
    0 & \mu_{2} & 0 & \mu_{3} \\
    0 & 0 & \mu_{4} & 0 \\
    0 & \mu_{3} & 0 & \mu_{5}
  \end{bmatrix}
  \begin{bmatrix}
    \bep^{\uu1} \\ \bep^{\uu5} \\ \bep^{\ww4} \\ \bep^{\ww5}
  \end{bmatrix}.
\end{equation}
The pure phonon elastic constants $\mu_{1},\mu_{2}$ are related to the
Lam\'e-constants $\mu,\lambda$ by
\begin{equation}
  \label{eq:mu1-mu2-lambda-mu}
  \mu_{1} = 2\mu + 3\lambda  \qquad  \mu_{2} = 2\mu,
\end{equation}
expressing the fact that an icosahedral quasicrystal behaves like an
isotropic medium when phason dynamics are frozen out.

$\mu_{4}$ and $\mu_{5}$ describe the elastic energy of pure phason
strain.  Contrary to the phonon case above, the phason
elasticity is not isotropic, when these two constants are different.
To simplify the calculations we assume in a first step that 
\begin{equation}
  \label{eq:spherical-approximation}
  \mu_{4}=\mu_{5} =:K_{1}
\end{equation}
is valid.  We have made this assumption, because the $SO(3)$
irreducible representation $\RP^{l=4}$ divides under icosahedral group
operations into $\RP^{4} + \RP^{5}$, which corresponds to the
splitting of the phason strain field $\supw{\bep}$ (see eq.
\eqref{eq:trafo-strain}).  If one demands not only icosahedral but
spherical symmetry in the space of phason strains, the identity of the
two phasonic elastic constants is necessary.  Therefore we call
eq. \eqref{eq:spherical-approximation} \emph{spherical approximation}.

Phonon and phason elasticity are coupled by $\mu_{3}$.  The assumption
that this coupling is small,
\begin{equation}
  \label{eq:small-mu3}
  \mu_{3} \ll \mu_{1},\mu_{2},\mu_{4}=\mu_{5},
\end{equation}
allows us to develop a perturbation theory with respect to $\mu_{3}$.

In ideal materials (without dislocations or defects) the only sources
for stress fields are external volume forces $\bbb$. Hence, the balance
of forces is expressed by
\begin{equation}
  \label{eq:balance-stresses-forces}
  \Div\bsi + \bbb = 0 .
\end{equation}
Please note, that the external force $\bbb=[\supu{\bbb},\supw{\bbb}]$,
as the displacement $[\buu,\bww]$, consists of two parts---the usual
phonon force $\supu{\bbb}$ and the so-called phason force
$\supw{\bbb}$.  Inserting Hooke's law
\eqref{eq:Hooke-law} in the balance of stresses and forces
\eqref{eq:balance-stresses-forces} and applying the definition
\eqref{eq:definition-strain} of the strain leads to generalized
elastic equations:
\begin{equation}
  \label{eq:displacement-equation}
  \bDD(\nabla)
  \begin{bmatrix}
    \buu\\\bww
  \end{bmatrix}
  + \bbb = 0 .
\end{equation}
This system of six second order partial differential equations
describes the connection between applied forces and displacement
field.  In this paper we provide an approximative solution by the
method of Green's function.

The differential operator $\bDD(\nabla)$ is a $6\times6$-matrix and can
be separated into four $3\times3$-matrices:
\begin{equation}
  \label{eq:diffop-four-diffops}
  \bDD(\nabla) = 
  \begin{bmatrix}
    \bDD^{\uu,\uu}(\nabla) & \bDD^{\uu,\ww}(\nabla) \\
    \bDD^{\ww,\uu}(\nabla) & \bDD^{\ww,\ww}(\nabla)
  \end{bmatrix}.
\end{equation}
For example, the second component of the phonon force is related to
the displacements by
$\DD^{\uu,\uu}_{2i}\uu_{i}+\DD^{\uu,\ww}_{2i}\ww_{i}+\bb_{2}=0$.  The
phonon-phonon block $\bDD^{\uu,\uu}$ is the well known differential
operator for the isotropic elastic continuum:
\begin{equation}
  \label{eq:diffop-isotropic}
  \bDD^{\uu,\uu}(\nabla) = \mu\Lapl + (\mu+\lambda)\Grad\Div.
\end{equation}
Applying the spherical approximation for the phason-pha\-son-block
explained above yields
\begin{equation}
  \label{eq:diffop-ww-spherical}
  \bDD^{\ww,\ww}(\nabla) = K_{1}\Lapl.
\end{equation}
The phonon-phason coupling is described by
\begin{multline}
  \label{eq:diffop-uw}
  \bDD^{\ww,\uu}(\nabla) = (\bDD^{\uu,\ww}(\nabla))^{t} \\
  = 
  \frac{\mu_{3}}{\sqrt{6}}
  \begin{bmatrix}
    F_{1}(x,y,z) & F_{3}(x,y) & F_{2}(z,x) \\
    F_{2}(x,y) & F_{1}(y,z,x) & F_{3}(y,z) \\
    F_{3}(z,x) & F_{2}(y,z) & F_{1}(x,y,z)
  \end{bmatrix}.
\end{multline}
with
\begin{equation}
 \begin{aligned}
  F_{1}(a,b,c) &= 
    -\frac{\partial^{2}}{\partial{a}^{2}}
    -\frac{1}{\tau}\frac{\partial^{2}}{\partial{b}^{2}}
    +\tau\frac{\partial^{2}}{\partial{c}^{2}} \\
  F_{2}(a,b) &=
    -2\frac{1}{\tau}\frac{\partial^{2}}{\partial{a}\partial{b}} \\
  F_{3}(a,b) &= 2\tau\frac{\partial^{2}}{\partial{a}\partial{b}} .
 \end{aligned}
\end{equation}

%

%
\subsection{Method of Green's function} 
\label{elastic-theory-greens-function}
The Green's function method is commonly used to solve linear inhomogenous
differential equations like \eqref{eq:displacement-equation}.
Expressing the displacement field $[\buu,\bww](\bxx)$ as a linear
combination of the applied forces $\bbb$,
\begin{equation}
  \label{eq:green-ansatz}
    \begin{bmatrix}
      \buu\\\bww
    \end{bmatrix}(\bxx)
    = \int\bGR(\bxx-\bxx')\bbb(\bxx')\dd^{3}\xx',
\end{equation}
leads to a system of differential equations for the \emph{Green's
  function} $\bGR(\bxx)$:
\begin{equation}
  \label{eq:diff-eq-greens-function}
  \bDD(\nabla)\bGR(\bxx) + \Eins\delt(\bxx) = 0.
\end{equation}
The operator $\bDD$ and the Green's function $\bGR$ are
$6\times6$-matrices, $\Eins$ is the six\dimensional{} identity operator
and $\delt(\bxx)$ the three\dimensional{} delta function.

By Fourier methods this system of differential equations can be
transformed into a solvable system of algebraic
equations\cite{itapdb:Ding1995a}.  The problem is that until now a
method for Fourier back transformation has not been provided.

%
%
\subsection{Solution for spherical approximation and noninteraction}
\label{elastic-theory-solution-noninteraction}

If we assume that phonon elasticity is not coupled with phason
elasticity ($\mu_{3}=0$), the differential equations
\eqref{eq:diff-eq-greens-function} become very simple:
\begin{align}
  \label{eq:diffeq-noninteracting-phonon}
  & \bDD^{\uu,\uu}(\nabla)\bGR_{0}^{\uu,\uu}(\bxx) + \Eins\delt(\bxx) =0 \\
  \label{eq:diffeq-noninteracting-phason}
  & \bDD^{\ww,\ww}(\nabla)\bGR_{0}^{\ww,\ww}(\bxx) + \Eins\delt(\bxx) =0.
\end{align}
Now $\Eins$ is the three\dimensional{} identity operator and the Green's
function matrix $\bGR$ is divided into four blocks by analogy with the
differential operator $\bDD$ in \eqref{eq:diffop-four-diffops}.  The
solutions of these two systems are given by
\begin{align}
  \label{eq:green-solution-noninteracting-phonon}
  \bGR_{0}^{\uu,\uu}(\bxx) &= -\frac{1}{8\pi\mu(\lambda+2\mu)}
    \{
      (\lambda+3\mu)\frac{\Eins}{\abs{\bxx}} 
      + (\lambda+\mu)\frac{\bxx\otimes\bxx}{\abs{\bxx}^{3}}
    \} \\
  \label{eq:green-solution-noninteracting-phason}
  \bGR_{0}^{\ww,\ww}(\bxx) &= \frac{1}{4\pi K_{1}}\Eins\frac{1}{\abs{\bxx}},
\end{align}
where $\bxx\otimes\bxx$ denotes the dyadic product of $\bxx$ with
itself.  \textname{Lord Kelvin} \cite{itapdb:Thomson1848} has provided
solution \eqref{eq:green-solution-noninteracting-phonon}. Eq.
\eqref{eq:green-solution-noninteracting-phason} is mathematically
equivalent to the solution for the electric potential in a given
charge density (see for example \textname{Feynman} \cite[II
6-1]{itapdb:Feynman1963}). 

In the next section we develop a perturbation theory with respect to
$\mu_{3}$, where $\bGR_{0}$ from
\eqref{eq:green-solution-noninteracting-phonon},
\eqref{eq:green-solution-noninteracting-phason} are the solutions of
zeroth order.

%

%

%
\section{Solution for finite phonon-phason interactions} 
\label{solution}

Based on the solution for vanishing coupling constant ($\mu_{3}=0$)
given in the previous section we now develop an approximative solution
for small but finite coupling between phonon and phason elasticity.
Section \ref{solution-perturbation} provides the perturbation of the
Green's function $\bGR(\bxx)$ with respect to $\mu_{3}$ and yields some
relations for $\bGR(\bxx)$.  In section \ref{solution-invariants} we
construct an invariant basis for the perturbative solutions, which
fulfills these relations.  Finally, in section
\ref{solution-coefficients} we find the solutions up to the order $2$
in $\mu_{3}$.

\subsection{Perturbation theory} 
\label{solution-perturbation}
The differential operator $\bDD(\nabla)$ defined in
\eqref{eq:diffop-four-diffops} consists of two parts
\begin{equation}
  \label{eq:diffop-perturbation}
  \bDD(\nabla) = \bDD_{0}(\nabla) + \bDD_{1}(\nabla)
\end{equation}
with
\begin{align}
  \label{eq:diffop-D0-def}
  \bDD_{0}(\nabla) &= 
  \begin{bmatrix}
    \bDD^{\uu,\uu}(\nabla) & 0 \\
    0  & \bDD^{\ww,\ww}(\nabla)
  \end{bmatrix}
  \\\label{eq:diffop-D1-def}
  \bDD_{1}(\nabla) &= 
  \begin{bmatrix}
    0 & \bDD^{\uu,\ww}(\nabla) \\
    \bDD^{\ww,\uu}(\nabla) & 0
  \end{bmatrix},
\end{align}
where the subscripts $0,1$ denote the order of $\mu_{3}$ (see
\eqref{eq:diffop-uw}).  Inserting \eqref{eq:diffop-perturbation} and
the expansion of the Green's function $\bGR$
\begin{equation}
  \label{eq:green-perturbation-ansatz}
  \bGR(\bxx) = \bGR_{0}(\bxx) + \bGR_{1}(\bxx) + \bGR_{2}(\bxx) + \ldots
\end{equation}
into the system of differential equations
\eqref{eq:diff-eq-greens-function} and comparing the terms with the
same order in $\mu_{3}$ leads to
\begin{eqnarray}
  \label{eq:diffeq-order-0}
  \bDD_{0}(\nabla)\bGR_{0}(\bxx) + \Eins\delt(\bxx) &=& 0 \\
  \label{eq:diffeq-order-n}
  \bDD_{0}(\nabla)\bGR_{n}(\bxx) + \bDD_{1}(\nabla)\bGR_{n-1}(\bxx) &=& 0
    \quad n=1,2,\ldots
\end{eqnarray}
From the form \eqref{eq:diffop-D0-def} and \eqref{eq:diffop-D1-def} of
$\bDD_{0}$ and $\bDD_{1}$ it follows immediately that the
phonon-phason-coupling blocks of the Green's function of even order
vanish, while the only remaining ones are of odd order:
\begin{equation}
  \label{eq:green-vanish}
  \bGR_{2i}^{\uu,\ww} = \bGR_{2i}^{\ww,\uu} =
  \bGR_{2i+1}^{\uu,\uu} = \bGR_{2i+1}^{\ww,\ww} = 0 \quad
  i=0,1,\ldots
\end{equation}
Because equation \eqref{eq:diffeq-order-0} of zeroth order is equivalent
to the noninteracting equations
\eqref{eq:diffeq-noninteracting-phonon}
\eqref{eq:diffeq-noninteracting-phason}, the nonvanishing part of the
Green's function $\bGR_{0}$ is given by
\eqref{eq:green-solution-noninteracting-phonon} and
\eqref{eq:green-solution-noninteracting-phason}.  It is necessary to
solve \eqref{eq:diffeq-order-n} to obtain the Green's functions of higher
order.

Let us now consider the analytic form of the solutions.  Four
properties result from the system of differential
equations \eqref{eq:diff-eq-greens-function}:
\begin{itemize}
\item \emph{Homogeneity} of degree $-1$
  \begin{equation}
    \label{eq:homogeneous-def}
    \bGR(\alpha\bxx) = \alpha^{-1}\bGR(\bxx) \qquad\alpha\in\RE 
  \end{equation}
\item \emph{Inversion symmetry}
  \begin{equation}
    \label{eq:invers-symmetry}
    \bGR(-\bxx) = \bGR(\bxx)
  \end{equation}
\item \emph{Transposing symmetry}
  \begin{equation}
    \label{eq:transpose-symmetry}
    \GR_{ij}(\bxx) = \GR_{ji}(\bxx)
  \end{equation}
\item \emph{Icosahedral symmetry}
  \begin{equation}
    \label{eq:icosahedral-symmetry}
    \bGR(\bxx) = \RP^{6}(g)\bGR(\RP^{3}(g)^{-1}\bxx)\RP^{6}(g)^{-1}
  \end{equation}
\end{itemize}
The six\dimensional{} representation $\RP^{6}$ describes the
transformation of the six\dimensional{} vector $[\buu,\bww]$.  It is the
sum of the two irreducible representations $\RP^{3}$ and $\RP^{3'}$.
Each part $\bGR_{n}(\bxx)$ of the perturbation series must have these
four properties, too.  The relations \eqref{eq:homogeneous-def} and
\eqref{eq:invers-symmetry} are valid for each matrix element
$\bGR_{n;i,j}(\bxx)$.  Especially the matrix elements of the solution
of zeroth order can be written as
\begin{equation}
  \label{eq:form-solution-zero-order}
  \GR_{0;ij}(\bxx) =\frac{1}{\abs{\bxx}} 
    \sum_{\substack{l=0\\l\textup{ even}}}^{l_{\textup{max}}=2}
    \sum_{m=-l}^{l} g_{0;ij}^{l,m} Y_{l}^{m}(\bxx),
\end{equation}
where $Y_{l}^{m}(\bxx)$ are the \emph{spherical harmonics} in cartesian
coordinates. 

The solutions of higher order can be expressed in the same form with
higher $l_{\textup{max}}$.  It is possible to make an ansatz with
coefficients $g_{n;i,j}^{l,m}$ and to transform
\eqref{eq:diffeq-order-n} into an algebraic system of equations. But
to avoid an exploding number of coefficients, we take into account that
each solution $\bGR_{n}(\bxx)$ must be invariant under icosahedral
transformations \eqref{eq:icosahedral-symmetry}.

The separation of the Green's function $\bGR(\bxx)$ into the four blocks
$\bGR^{\alpha,\beta}(\bxx)$ with $\alpha,\beta\in\{\uu,\ww\}$
transforms relations \eqref{eq:transpose-symmetry} and
\eqref{eq:icosahedral-symmetry} into
\begin{equation}
  \label{eq:transpose-symmetry-blocks}
  \GR^{\uu,\uu}_{ij} = \GR^{\uu,\uu}_{ji} \qquad
  \GR^{\ww,\ww}_{ij} = \GR^{\ww,\ww}_{ji} \qquad
  \GR^{\ww,\uu}_{ij} = \GR^{\uu,\ww}_{ji}
\end{equation}
and
\begin{equation}
  \label{eq:icosahedral-symmetry-blocks}
  \bGR^{\alpha,\beta}(\bxx) = 
  \RP^{\alpha}(g)\bGR^{\alpha,\beta}(\RP^{3}(g)^{-1}\bxx)\RP^{\beta}(g)^{-1}  
\end{equation}
with the definition $\RP^{\uu}:=\RP^{3}$ and $\RP^{\ww}:=\RP^{3'}$.
In the next section we construct icosahedral invariants for the four
blocks, e.g. $3\times3$-matrix-functions, which fulfill relations
\eqref{eq:homogeneous-def}, \eqref{eq:invers-symmetry},
\eqref{eq:transpose-symmetry-blocks} and
\eqref{eq:icosahedral-symmetry-blocks}. 

%
%

%
\subsection{Icosahedral invariants} 
\label{solution-invariants}
Any $3\times3$-matrix-function can be expressed as
\begin{equation}
  \label{eq:matrix-function}
  \sum_{i}M_{i}\ff_{i}(\bxx)
\end{equation}
with $3\times3$-basis-matrices $M_{i}$ and scalar functions
$\ff_{i}(\bxx)$.  To achieve a basis in the space of icosahedral
invariant matrix-functions it is necessary to divide the space of
matrices and the space of scalar functions into icosahedral invariant
subspaces, i.e. to construct the icosahedral irreducible bases (see
for example \cite{itapdb:Trebin1979}).  By
building the inner products of all possible combinations of
irreducible bases of the matrices and the functions, where both bases
belong to the same irreducible representation, we obtain a complete
basis of the space of icosahedral invariant matrix-functions:
\begin{equation}
  \label{eq:invariant-basis-definition}
  I^{\alpha,\beta}_{n}(\bxx) = \frac{1}{\sqrt{d_{\gamma}}}
  \sum_{i=1}^{d_{\gamma}} M_{i}^{\alpha,\beta;\gamma}
  \ff_{i}^{l;\gamma}(\bxx).
\end{equation}
In this formula $\alpha,\beta\in\{3,3'\}$ specify the block and
$\gamma$ denotes the irreducible representation $\RP^{\gamma}$ of
dimension $d_{\gamma}$, to which the irreducible bases
$\{M_{i}^{\alpha,\beta;\gamma}\}_{i=1\ldots{}d_{\gamma}}$ and
$\{\ff_{i}^{l;\gamma}(\bxx)\}_{i=1\ldots{}d_{\gamma}}$ belong.  The
index $n$ on the left hand side is a function of the representation
index $\gamma$ and the orbital quantum number $l$ on the right.  It is
the enumeration of all possible $(\gamma,l)$ combinations, ordered
first by $l\in\{0,2,4,6,8\}$ and second by $\gamma\in\{1,4,5\}$.  In
the next two paragraphs we discuss the two irreducible bases.

The irreducible basis of the $3\times3$-matrices $M$ can easily be
calculated for all four blocks, but because of
\eqref{eq:transpose-symmetry-blocks} it is sufficient to consider the
three cases $(\uu,\uu)$, $(\ww,\ww)$ and $(\ww,\uu)$.  The
\emph{Clebsch-Gordan-Series} are given by
\begin{itemize}
\item $\RP^{3}\otimes\RP^{3}=\RP^{1}\oplus\RP^{3}\oplus\RP^{5}$
\item $\RP^{3'}\otimes\RP^{3}=\RP^{4}\oplus\RP^{5}$
\item $\RP^{3'}\otimes\RP^{3'}=\RP^{1}\oplus\RP^{3'}\oplus\RP^{5}$.
\end{itemize}
The explicit form of the irreducible basis matrices is given in
appendix \ref{basis-matrix}.  The matrices which belong to the
irreducible representations $\RP^{3}$ and $\RP^{3'}$ are
antisymmetric.  We do not need them for our purpose, because the
Green's function must be symmetric (\ref{eq:transpose-symmetry},
\ref{eq:transpose-symmetry-blocks}).

The space of homogeneous functions of degree $0$ consists of spherical
invariant subspaces, denoted by the orbital quantum number $l$, whose
bases are given by $B_{l}=\{Y_{l}^{m}(\bxx)\}_{m=-l\ldots{}l}$.  The
inversion symmetry \eqref{eq:invers-symmetry} allows us to limit the
considerations to even quantum numbers $l$.  The icosahedral
group is a subgroup of the spherical group $SO(3)$. The spherical
irreducible basis $B_{l}$ is icosahedrally reducible for each $l>2$.
We have calculated the icosahedral irreducible bases
$\{\ff_{i}^{l;\gamma}(\bxx)\}_{i\ldots{}d_{\gamma}}$ from $B_{l}$ for
$l=0,2,4,6,8$.  They are given in appendix \ref{basis-function} for
$l=0,2$.

Combining all possible pairs of irreducible matrices and irreducible
functions using equation \eqref{eq:invariant-basis-definition} leads
to a set of icosahedral invariant matrix functions
$I_{n}^{\alpha,\beta}(\bxx)$, which shows inversion
\eqref{eq:invers-symmetry} and transposing
\eqref{eq:transpose-symmetry-blocks} symmetry.  These invariants are
homogeneous of degree $0$.  By dividing them by $\abs{\bxx}$ one
obtains homogeneity of degree $-1$ \eqref{eq:homogeneous-def}.

%
%

%
\subsection{Solution in terms of coefficients} 
\label{solution-coefficients}
It is possible to express the Green's function of order $0$
in terms of the icosahedral invariant matrix functions calculated in
the previous section.  Inserting
\begin{align}
  \label{eq:1-xx-invariant}
  \Eins &= \sqrt{3}I^{3,3}_{1}(\bxx) = \sqrt{3}I^{3',3'}_{1}(\bxx)
  \\
  \frac{\bxx\otimes\bxx}{\bxx^{2}} &= 
  \frac{1}{\sqrt{3}}I^{3,3}_{1}(\bxx) + \sqrt{5}I^{3,3}_{2}(\bxx) 
\end{align}
in equations \eqref{eq:green-solution-noninteracting-phonon} and
\eqref{eq:green-solution-noninteracting-phason} leads to
\begin{multline}
  \bGR_{0}^{\uu,\uu}(\bxx) =
  -\frac{1}{24\pi\mu(\lambda+2\mu)\abs{\bxx}}
  \\
  \{2\sqrt{3}(2\lambda+5\mu)I_{1}^{3,3}(\bxx)
  + 3\sqrt{5}(\lambda+\mu)I_{2}^{3,3}(\bxx)\}
\end{multline}
\begin{equation}
  \bGR_{0}^{\ww,\ww}(\bxx) = \frac{\sqrt{3}}{4\pi{}K_{1}\abs{\bxx}}
  I_{1}^{3',3'}(\bxx).
\end{equation}
The Green's functions of order $1$ and $2$ are calculated by inserting
an ansatz in the invariant matrix functions into the recursion
equation \eqref{eq:diffeq-order-n} and solving the systems of
algebraic equations in the coefficients.  The solutions are:
  \begin{multline}
    \label{eq:green-order-1}
    \bGR_{1}^{\ww,\uu}(\bxx) = 
    \frac{\mu_{3}}{224\pi{}K_{1}\mu(\lambda+2\mu)\abs{\bxx}}
    \\
    \{
    4\sqrt{5}(4\lambda+11\mu)I_{1} +
    7\sqrt{15}(\lambda+\mu)I_{2} +
    \\
    10\sqrt{21}(\lambda+\mu)I_{3}
    \}
  \end{multline}
  \begin{multline}
    \bGR_{2}^{\uu,\uu}(\bxx) =
    \frac{\mu_{3}^{2}}{\pi{}K_{1}\mu^{2}(\lambda+2\mu)^{2}\abs{\bxx}}
    \\
    \{ -\frac{\sqrt{3}}{126}(4\lambda^{2}+16\lambda\mu+19\mu^{2})I_{1}
    -\frac{\sqrt{5}}{84}(2\lambda^{2}+8\lambda\mu+5\mu^{2})I_{2}
    \\
    -\frac{5\sqrt{21}}{16016}(43\lambda^{2}+242\lambda\mu+342\mu^{2})I_{3}
    +\frac{25\sqrt{7}}{672}(\lambda+\mu)(\lambda+3\mu)I_{4}
    \\
    -\frac{5\sqrt{55}}{1056}(\lambda+\mu)(3\lambda+8\mu)I_{5}
    +\frac{15\sqrt{1186185}}{262912}(\lambda+\mu)^{2}I_{6}
    \\
    -\frac{35\sqrt{1185}}{7584}(\lambda+\mu)^{2}I_{7} \}
  \end{multline}
  \begin{multline}
    \bGR_{2}^{\ww,\ww}(\bxx) =
    \frac{\mu_{3}^{2}}{\pi{}K_{1}^{2}\mu(\lambda+2\mu)\abs{\bxx}} 
    \\
    \{
    -\frac{\sqrt{3}}{126}(4\lambda+11\mu)I_{1}
    -\frac{\sqrt{5}}{42}(\lambda+3\mu)I_{2} 
    +\frac{5\sqrt{21}}{2464}(2\lambda+13\mu)I_{3}
    \\
    +\frac{25\sqrt{7}}{672}(\lambda+\mu)I_{4}
    +\frac{25\sqrt{55}}{1056}(\lambda+\mu)I_{5} \}.
  \end{multline}
$I_{n}$ is an abbreviation for $I_{n}^{\alpha,\beta}(\bxx)$ with
appropriate $\alpha,\beta\in\{\uu,\ww\}$.

%
%

%
%
\section{Results and discussion} 
\label{results}
With the help of the Green's function calculated in the previous
section it is now possible to examine the displacement fields
$\buu(\bxx)$ and $\bww(\bxx)$ due to a force field.  Inserting the
point force
\begin{equation}
  \label{eq:force-field}
  \bbb(\bxx) = \bbb_{0}\delt(\bxx)
\end{equation}
into equation \eqref{eq:green-ansatz} leads to the displacement field
\begin{equation}
  \label{eq:displacement-point-force}
    \begin{bmatrix}
      \buu\\\bww
    \end{bmatrix}(\bxx)
    = \bGR(\bxx)\bbb_{0}.
\end{equation}
Because of the homogeneity of degree $-1$ \eqref{eq:homogeneous-def}
the radial behaviour is given by
\begin{equation}
  \label{eq:displacement-radial}
    \begin{bmatrix}
      \buu\\\bww
    \end{bmatrix}(\bxx)
    \propto \frac{1}{\abs{\bxx}}.
\end{equation}
It is more interesting to examine the variation of the displacement
fields on a sphere---the icosahedral symmetry should be reflected.
Let us now consider a phonon force $\supu{\bbb_{0}}$.  The
corresponding displacement fields in perturbation theory of second
order
\begin{align}
  \label{eq:displacement-phonon-force}
  \buu(\bxx) &= \bGR^{\uu,\uu}(\bxx)\supu{\bbb_{0}} =
    \bGR_{0}^{\uu,\uu}(\bxx)\supu{\bbb_{0}} 
    + \bGR_{2}^{\uu,\uu}(\bxx)\supu{\bbb_{0}} + \ldots \\
  \bww(\bxx) &= \bGR^{\ww,\uu}(\bxx)\supu{\bbb_{0}} =
    \bGR_{1}^{\ww,\uu}(\bxx)\supu{\bbb_{0}} + \ldots
\end{align}
show, that the \emph{isotropic} phonon displacement
$\bGR_{0}^{\uu,\uu}(\bxx)\supu{\bbb_{0}}$ is corrected by a term of second
order and that the phason displacement is given by a term of first
order.  We choose a spherical coordinate system with
$\phi\in[0\ldots{}2\pi]$ and $\theta\in[0\ldots\pi]$, which depends on
the force direction $\supu{\bbb_{0}}$ in such a way, that $\supu{\bbb_{0}}$
always points to the \emph{North Pole} ($\theta=0$).

This force breaks the full icosahedral symmetry, but nevertheless the
symmetry on parallels of latitude ($\theta=\textup{const}$) is
preserved: $n$-fold symmetry, when the force points into such a
direction of the icosahedron.  The figures show the absolute value of
the displacement fields $\uu$ (figures \ref{fig:u-2-abs-2-fold} and
\ref{fig:u-2-abs-5-fold}) and $\ww$ (figure \ref{fig:w-1-abs-3-fold})
on such a parallel with $\theta=\frac{\pi}{4}$.  For these figures we
have choosen the elastic constants as given by \textname{Jari\'c} and
\textname{Mohanty} \cite{itapdb:Jaric1988g}.

All figures reflect the deviation from the spherical symmetry but show
the icosahedral symmetry: the two\fold{} axis in figure
\ref{fig:u-2-abs-2-fold}, five\fold{} in figure
\ref{fig:u-2-abs-5-fold} and three\fold{} in figure
\ref{fig:w-1-abs-3-fold}.  In the case of phonon displacement this
deviation is less than one percent.  It is possible to examine the
three components of the displacement fields or different values of
$\theta$, but the results are very similar.

It is important to note that the deviation from the spherical symmetry
is present although we have made the spherical approximation in the
space of phason strains.  Because the space of phonon strains is
isotropic by definition, this deviation is only due to the nonvashing
coupling between these two spaces, depending in second order on
$\mu_{3}$. 

We will now consider the five\fold{} axis (figure
\ref{fig:u-2-abs-5-fold}) in detail.  The absolute value of the phonon
displacement field in second order on the parallel of latitude with
$\theta=\frac{\pi}{4}$ is given by
\begin{multline}
  \label{eq:displacement-phonon-second-absolute-parallel}
  \abs{\buu^{(2)}(\phi)} := 
  \abs{\bGR_{2}^{\uu,\uu}(r=1,\theta=\frac{\pi}{4},\phi)\supu{\bbb_{0}}} \\
  = \frac{\mu_{3}^{2}10^{-5}}{\pi{}K_{1}\mu^{2}}
  \sqrt{
    a_{0}(\frac{\lambda}{\mu})
   +a_{1}(\frac{\lambda}{\mu}) \cos(5\phi)
   +a_{2}(\frac{\lambda}{\mu}) \cos(10\phi)
  },
\end{multline}
where the coefficients $a_{0}$, $a_{1}$, and $a_{2}$ are functions in the
ratio $\lambda/\mu$.  They are plotted in figure
\ref{fig:a-nu-log} as a function of Poisson's ratio $\nu$, which is
connected to $\lambda$ and $\mu$ by
\begin{equation}
  \label{eq:lambda-mu-nu}
  \frac{\lambda}{\mu} = \frac{2\nu}{1-2\nu}.
\end{equation}
In the physically interesting interval $\nu\in[0,0.5]$ the coefficient
$a_{0}$ is half an order of magnitude larger than $a_{1}$, which again
is one
order of magnitude larger than $a_{2}$.  Therefore the maximum and the
minimum of $\abs{\buu^{(2)}(\phi)}$ are at the same positions as the
maximum and the minimum of $\cos(5\phi)$.  The difference between
maximum and minimum is then given by
\begin{equation}
  \label{eq:max-min-u-abs}
  \abs{\buu^{(2)}(0)}-\abs{\buu^{(2)}(\frac{\pi}{5})}.
\end{equation}

In this paper we have discussed the theory of elasticity in
icosahedral quasicrystals.  We provide for the first time an
approximative solution for the elastic Green's function, which
reflects the deviation of the isotropy.  For this approximation we
made the assumption, that 
\begin{itemize}
\item the two phason elastic constants are equal ($\mu_{4}=\mu_{5}$)
  and that
\item the coupling constant $\mu_{3}$ is small.
\end{itemize}
Measurements of \textname{Boudard, de Boissieu} et al
\cite{itapdb:Boudard1995a,itapdb:Boissieu1995} in AlPdMn show that the
first assumption is not fulfilled.  The constant $K_{2}$ should be
zero (see appendix \ref{constants}).  Nevertheless, the qualitative
results of this paper remain valid, because they are based on the
coupling between phonons and phasons.  The component of the phason
strain $\bep^{\ww5}$, which is coupled to the phonon strain, is
directly connected to the constant $\mu_{5}$.  The value of the other
phason constant $\mu_{4}$ does only indirectly affect this coupling.

The elastic Green's function is provided up to second order of
perturbation theory in a basis of icosahedral invariants.  With the
help of this Green's function we examined the phonon and phason
displacement fields due to a phonon point force.  Although the
icosahedral symmetry is close to the isotropic symmetry, the deviation
from the isotropy is observable.

\appendix
\section{Icosahedral basis} 
\label{basis}
\subsection{Icosahedral irreducible strain} 
\label{basis-strain}
We have choosen the same coordinate system as in
\cite{itapdb:Ishii1989e,itapdb:Trebin1991,itapdb:Litvin1991}.  Using
the golden number $\tau=(1+\sqrt{5})/2$ the icosahedral irreducible
strain is given by:
  \begin{align}
    \bep^{\uu1} &= \frac{1}{\sqrt{3}}
        (\supu{\ep}_{11}+\supu{\ep}_{22}+\supu{\ep}_{33}) \\
    \bep^{\ww4} &= \frac{1}{\sqrt{3}}
    \begin{bmatrix}
      \supw{\ep}_{11}+\supw{\ep}_{22}+\supw{\ep}_{33} \\
      \frac{1}{\tau}\supw{\ep}_{21}+\tau\supw{\ep}_{12} \\
      \frac{1}{\tau}\supw{\ep}_{32}+\tau\supw{\ep}_{23} \\
      \frac{1}{\tau}\supw{\ep}_{13}+\tau\supw{\ep}_{32}
    \end{bmatrix}  \\
    \bep^{\uu5} &= 
    \begin{bmatrix}
      \frac{1}{2\sqrt{3}}
      (-\tau^{2}\supu{\ep}_{11}+\frac{1}{\tau^{2}}\supu{\ep}_{22}
        +(\tau+\frac{1}{\tau})\supu{\ep}_{33}) \\
      \frac{1}{2}(\frac{1}{\tau}\supu{\ep}_{11}-\tau\supu{\ep}_{22}+\supu{\ep}_{33}) \\
      \sqrt{2}\supu{\ep}_{12} \\
      \sqrt{2}\supu{\ep}_{23} \\
      \sqrt{2}\supu{\ep}_{31}
    \end{bmatrix} \\
    \bep^{\ww5} &= \frac{1}{\sqrt{6}}
    \begin{bmatrix}
      \sqrt{3}(\supw{\ep}_{11}-\supw{\ep}_{22}) \\
      \supw{\ep}_{11}+\supw{\ep}_{22}-2\supw{\ep}_{33} \\
      \sqrt{2}(\tau\supw{\ep}_{21}-\frac{1}{\tau}\supw{\ep}_{12}) \\
      \sqrt{2}(\tau\supw{\ep}_{32}-\frac{1}{\tau}\supw{\ep}_{23}) \\
      \sqrt{2}(\tau\supw{\ep}_{13}-\frac{1}{\tau}\supw{\ep}_{31}) \\
    \end{bmatrix}.
  \end{align}
\subsection{Icosahedral irreducible matrices} 
\label{basis-matrix}
In order to write the icosahedral irreducible matrices in a compact
form it is useful to define a \emph{permutation operator} 
\begin{equation}
  \label{eq:permutation-operator-definition}
  P: M \mapsto P(M) \textup{ with } P(M)_{i,j}:=M_{i-1,j-1\mod 3},
\end{equation}
which permutes the matrix indices modulo 3.
\begin{itemize}
\item phonon-phonon-block:
  \begin{align*}
    M_{1}^{3,3;1} &= \frac{\sqrt{3}}{3}
    \begin{bmatrix}
      1&0&0\\0&1&0\\0&0&1
    \end{bmatrix}
    &
    M_{1}^{3,3;3} &= \frac{\sqrt{2}}{2}
    \begin{bmatrix}
      0&0&0\\0&0&1\\0&-1&0
    \end{bmatrix}
    \\
    M_{2}^{3,3;3} &= P(M_{1}^{3,3;3})
    & M_{3}^{3,3;3} &= P^{2}(M_{1}^{3,3;3})
\\
    M_{1}^{3,3;5} &= \frac{\sqrt{3}}{6}
    \begin{bmatrix}
      -\tau^{2}&0&0\\0&\frac{1}{\tau^{2}}&0\\0&0&\tau+\frac{1}{\tau}
    \end{bmatrix}
    \\
    M_{2}^{3,3;5} &= \frac{1}{2}
    \begin{bmatrix}
      \frac{1}{\tau}&0&0\\0&-\tau&0\\0&0&1
    \end{bmatrix}
    &
    M_{3}^{3,3;5} &= \frac{\sqrt{2}}{2}
    \begin{bmatrix}
      0&1&0\\1&0&0\\0&0&0
    \end{bmatrix}
    \\
    M_{4}^{3,3;5} &= P(M_{3}^{3,3;5})
    &
    M_{5}^{3,3;5} &= P^{2}(M_{3}^{3,3;5})
  \end{align*}
\item phason-phonon-block:
  \begin{align*}
    M_{1}^{3',3;4} &= M_{1}^{3,3;1}
    &
    M_{2}^{3',3;4} &= \frac{\sqrt{3}}{3}
    \begin{bmatrix}
      0&\tau&0\\\frac{1}{\tau}&0&0\\0&0&0
    \end{bmatrix}
    \\
    M_{3}^{3',3;4} &= P(M_{2}^{3',3;4})
    &
    M_{4}^{3',3;4} &= P^{2}(M_{2}^{3',3;4})
\\
    M_{1}^{3',3;5} &= \frac{\sqrt{2}}{2}
    \begin{bmatrix}
      -1&0&0\\0&1&0\\0&0&0
    \end{bmatrix}
    &
    M_{2}^{3',3;5} &= \frac{\sqrt{6}}{6}
    \begin{bmatrix}
      -1&0&0\\0&-1&0\\0&0&2
    \end{bmatrix}
    \\
    M_{3}^{3',3;5} &= \frac{\sqrt{3}}{3}
    \begin{bmatrix}
      0&\frac{1}{\tau}&0\\-\tau&0&0\\0&0&0
    \end{bmatrix}
    &
    M_{4}^{3',3;5} &= P(M_{3}^{3',3;5})
    \\
    M_{5}^{3',3;5} &= P^{2}(M_{3}^{3',3;5})
  \end{align*}
\item The icosahedral irreducible matrices of the phason-phason-block
  can be generated from the matrices of the phonon-phonon-block by the
  substitution 
  \begin{equation}
    \RP^{3}\rightarrow\RP^{3'}\qquad
    \tau\rightarrow-\frac{1}{\tau}.
  \end{equation}
\end{itemize}
All matrices are normalized with respect to the quadratic vector norm,
where $\{\delta_{ij}|i,j=1\ldots3\}$ is considered to be the
normalized basis of the vector space.

\subsection{Icosahedral irreducible functions} 
\label{basis-function}
\begin{equation}
  \ff_{1}^{0;1}(\bxx) = 1
\end{equation}
\begin{align}
  \ff_{1}^{2;5}(\bxx) &= \frac{\sqrt{3}}{6\bxx^{2}} 
  [-\tau^{2}x^{2}+\frac{1}{\tau^{2}}y^{2}+(\tau+\frac{1}{\tau})z^{2}]
  \\
  \ff_{2}^{2;5}(\bxx) &= \frac{1}{2\bxx^{2}} 
  [\frac{1}{\tau}x^{2}-\tau{}y^{2}+z^{2}]
\end{align}
\begin{align}
  \ff_{3}^{2;5}(\bxx) &= \frac{\sqrt{2}}{\bxx^{2}}xy
  &
  \ff_{4}^{2;5}(\bxx) &= \frac{\sqrt{2}}{\bxx^{2}}yz
  &
  \ff_{5}^{2;5}(\bxx) &= \frac{\sqrt{2}}{\bxx^{2}}zx
\end{align}
All functions are normalized with respect to the quadratic vector
norm, where
$\{\sqrt{\frac{i!j!k!}{l!}}\frac{x^{i}y^{j}z^{k}}{\bxx^{l}}%
|i,j,k\ge0,i+j+k=l\}$
is considered to be the normalized basis of the vector space, which is
the space of all spherical harmonics to the orbital quantum number
$l$.

\section{Elastic constants} 
\label{constants}
Unfortunately, the definitions of the elastic constants of icosahedral
quasicrystals differ in the literature.  Therefore we want to compare
our definition given in \eqref{eq:Hooke-law-explicit} to others.
  \begin{itemize}
  \item \textname{Trebin}, \textname{Fink},
    \textname{Stark}\cite{itapdb:Trebin1991,itapdb:Trebin1993c} 
    \begin{equation}
      \label{eq:ela-const-trebin}
      \begin{aligned}
        \mu_{1}&=3\lambda_{1}\qquad &
        \mu_{2}&=\lambda_{2}\qquad &
        \mu_{3}&=\lambda_{3}\qquad \\
        \mu_{4}&=\lambda_{4}\qquad &
        \mu_{5}&=\lambda_{5}\qquad 
      \end{aligned}
    \end{equation}
  \item \textname{Shaw}, \textname{Elser},
    \textname{Henley}\cite{itapdb:Shaw1991},
    \textname{Widom}\cite{itapdb:Widom1991a} and
    \textname{Henley}\cite{itapdb:Henley1991a} use the well known
    Lam\'e-constants $\lambda$ and $\mu$, see
    \eqref{eq:mu1-mu2-lambda-mu}.  The three remaining constants are:
    \begin{equation}
      \mu_{3} = K_{3}\sqrt{6} \qquad
      \mu_{4} = K_{1} + \frac{5}{3} K_{2} \qquad
      \mu_{5} = K_{1} - \frac{4}{3} K_{2} 
    \end{equation}
    \textname{Boudard}, \textname{de Boissieu} et al
    \cite{itapdb:Boudard1995a,itapdb:Boissieu1995} use the same
    elastic constants (see reference 19 in \cite{itapdb:Boissieu1995}).
  \item \textname{Ding et al} \cite{itapdb:Ding1993} also use the
    Lam\'e-constants, but the \emph{phasonic} constants are different:
    \begin{equation}
      \mu_{3} = R\sqrt{6} \qquad
      \mu_{4} = K_{1} - 2K_{2} \qquad
      \mu_{5} = K_{1} + K_{2}.
    \end{equation}
  \end{itemize}

\bibliographystyle{epj}
\bibliography{jourkurz,paper1}

\section*{Figures} 
%
\makeatletter
\def\@captype{figure}
\makeatother
%
\newcommand{\includefigure}[3]%
{\pagebreak[1]\centerline{\includegraphics[bb=0 0 #1 #2,width=8cm]{#3}}}
%
\includefigure{300}{200}{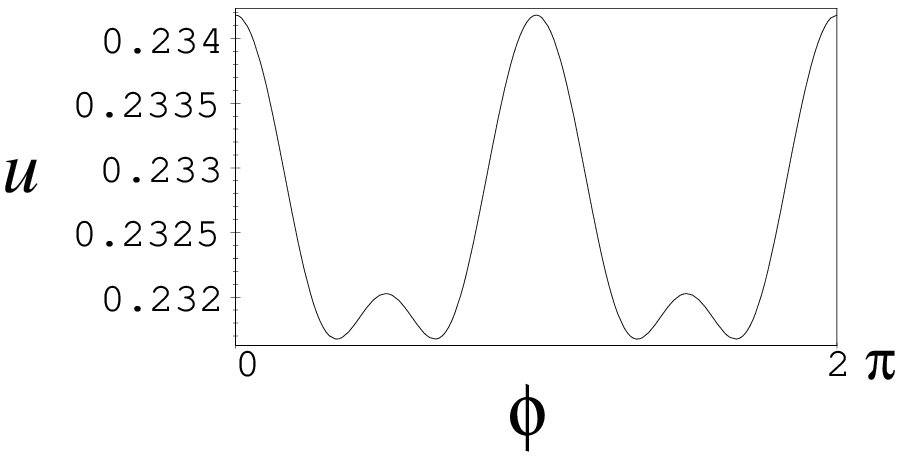}
\caption{Absolute value of phonon displacement $\abs{\uu}$ versus $\phi$
  for $\theta=\frac{\pi}{4}$, 2-fold axis.}
\label{fig:u-2-abs-2-fold}
\includefigure{300}{200}{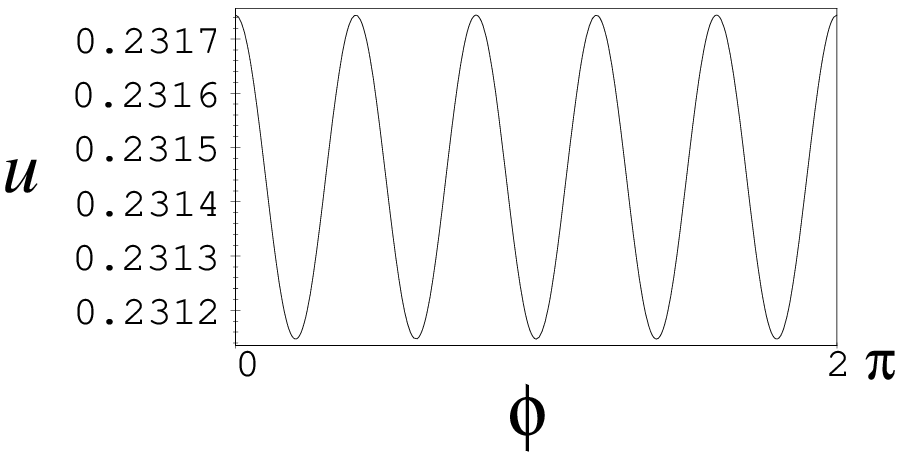}
\caption{Same as figure \ref{fig:u-2-abs-2-fold}, but 5-fold axis.}
\label{fig:u-2-abs-5-fold}
\includefigure{300}{200}{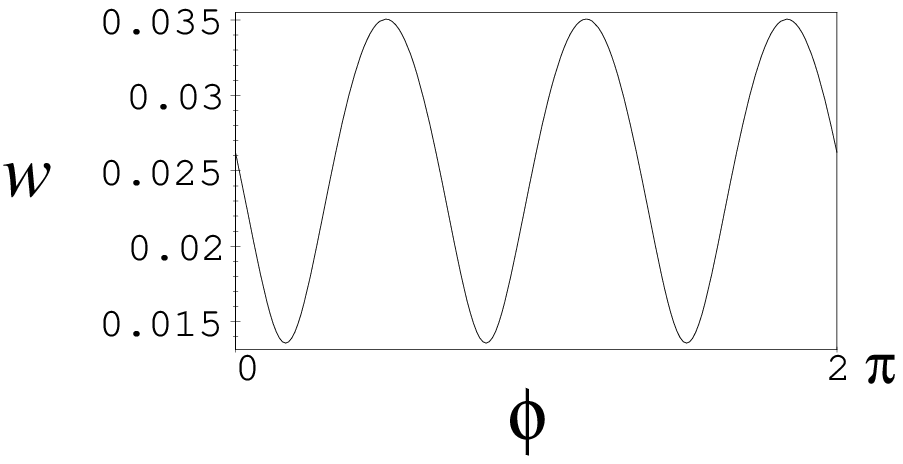}
\caption{Absolute value of phason displacement $\abs{\ww}$ versus $\phi$
  for $\theta=\frac{\pi}{4}$, 3-fold axis.}
\label{fig:w-1-abs-3-fold}
\includefigure{300}{250}{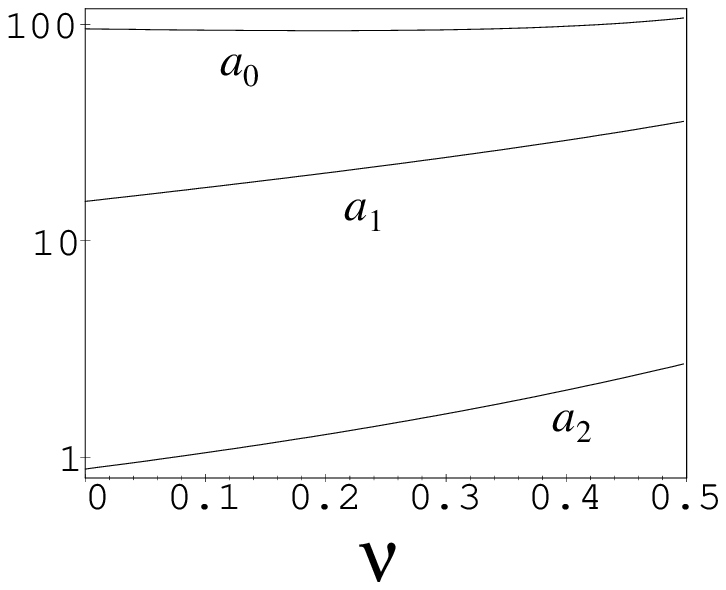}
\caption{Coefficients $a_{0},a_{1}$, and $a_{2}$ as a function of
  Poisson's ratio $\nu$.}
\label{fig:a-nu-log}

\end{document}